# STIED: A deep learning model for the SpatioTemporal detection of focal Interictal Epileptiform Discharges with MEG

Raquel Fernández-Martín*, Alfonso Gijón*, Odile Feys, Elodie Juvené, Alec Aeby, Charline Urbain,
Xavier De Tiège#, Vincent Wens#

*Abstract*—**Magnetoencephalography (MEG) allows the non-invasive detection of interictal epileptiform discharges (IEDs). Clinical MEG analysis in epileptic patients traditionally relies on the visual identification of IEDs, which is time consuming and partially subjective. Automatic, data-driven detection methods exist but show limited performance. Still, the rise of deep learning (DL)— with its ability to reproduce human-like abilities—could revolutionize clinical MEG practice. Here, we developed and validated STIED, a simple yet powerful supervised DL algorithm combining two convolutional neural networks with temporal (1D time-course) and spatial (2D topography) features of MEG signals inspired from current clinical guidelines. Our DL model enabled both temporal and spatial localization of IEDs in patients suffering from focal epilepsy with frequent and high amplitude spikes (FE group), with high-performance metrics—accuracy, specificity, and sensitivity all exceeding 85%—when learning from spatiotemporal features of IEDs. This performance can be attributed to our handling of input data, which mimics established clinical MEG practice. Reverse engineering further revealed that STIED encodes fine spatiotemporal features of IEDs rather than their mere amplitude. The model trained on the FE group also showed promising results when applied to a separate group of presurgical patients with different types of refractory focal epilepsy, though further work is needed to distinguish IEDs from physiological transients. This study paves the way of incorporating STIED and DL algorithms into the routine clinical MEG evaluation of epilepsy.**

## I. INTRODUCTION

Epilepsy is a disorder of the brain characterized by an enduring predisposition to generate epileptic seizure [1]. It affects people of all ages and has neurobiological, cognitive, psychological and social consequences. While pharmacological treatments work successfully in most patients, approximately one-third of them suffer from refractory (i.e., drug-resistant) epilepsy [2]. These patients are possible epilepsy surgery candidates and must therefore undergo a presurgical evaluation aiming at delineating as precisely as possible the epileptogenic zone.

Magnetoencephalography (MEG) contributes to this evaluation, mainly by allowing the non-invasive detection of interictal epileptiform discharges (IEDs)—subclinical events occurring between seizures [3], [4]—which are markers of the localization of the epileptogenic zone [5], [6]. Clinical MEG analysis of patients with epilepsy traditionally relies on the visual identification of IEDs by an experienced clinical magnetoencephalographer, who inspects both IED sensor time courses and spatial topographies along with their source localization usually by equivalent current dipole (ECD) modeling [7], [8]. This approach is highly time consuming (on average 8 hours per patient [9]) and partially subjective (notwithstanding existing guidelines [10]). The development of efficient data-driven, automatic approaches to IED detection would represent a major advance in clinical MEG practice. Several proposals based on unsupervised algorithms have been

This work was supported by the Fonds Erasme (Brussels, Belgium; Research Convention: « Les Voies du Savoir»). Odile Feys is supported by the Fonds pour la Formation à la Recherche dans l'Industrie et l'Agriculture (FRIA, Fonds de la Recherche Scientifique (FRS-FNRS), Brussels, Belgium). Xavier De Tiège is Clinical Researcher at the FRS-FNRS (Brussels, Belgium). The MEG project at the Hôpital Universitaire de Bruxelles is financially supported by the Fonds Erasme.

#Xavier De Tiège (X. D. T.) and Vincent Wens (V. W.) contributed equally to this work.
*Corresponding authors: Raquel Fernández-Martín (R. F. M.); Alfonso Gijón (A. G.).
*R. F. M. is with the Université libre de Bruxelles (ULB), ULB Neuroscience Institute (UNI), Laboratoire de Neuroanatomie et de Neuroimagerie translationnelles (LN²T), Brussels, Belgium. (e-mail: raquel.fernandez.martin@ulb.be).
*A. G. is with the Universidad de Córdoba, Department of Informatics and Numerical Analysis, Córdoba, Spain. (email: agijon@uco.es)
O. F. is with the ULB, UNI, LN²T and ULB, Hôpital Universitaire de Bruxelles (H.U.B), Hôpital Erasme, Department of Neurology, Brussels, Belgium.
E. J. is with the ULB, H.U.B, Department of Pediatric Neurology, Brussels, Belgium.
A. A. is with the ULB, H.U.B, Department of Pediatric Neurology, Brussels, Belgium.
C. U. is with the ULB, UNI, Center for Research in Cognition and Neurosciences (CRCN), Neuropsychology and Functional Neuroimaging Unit (UR2NF), Brussels, Belgium.
X. D. T. is with the ULB, UNI, LN²T, and ULB H.U.B., Hôpital Erasme, Service of translational Neuroimaging, Brussels, Belgium.
V. W. is with the ULB, UNI, LN²T, and ULB, H.U.B., Hôpital Erasme, Service of translational Neuroimaging, Brussels, Belgium.

explored, e.g., Independent Component Analysis (ICA) and Hidden Markov Modeling (HMM), with limitations regarding detection specificity or full automation [11]–[17].

In the last years, supervised deep learning (DL) algorithms emerged as novel automatic approaches [18]–[21]. These methods are mainly based on convolutional neural networks (CNNs) extracting temporal features of IED waveforms from MEG signals, and then supplemented with somewhat *ad-hoc* techniques for the subsequent spatial localization of epileptic events, e.g., splitting sensor channels into different brain regions [18], [21], classification and segmentation networks [19], or graph convolutional networks [20]. While promising, none of these approaches performed well enough to be incorporated into routine clinical MEG practice.

Here, we hypothesized that a reliable, high-performance DL algorithm should combine both temporal (spiking waveform) and spatial (dipolar magnetic pattern) features used by clinical magnetoencephalographers to visually identify and select epileptiform discharges in MEG signals [10]. For that purpose, we designed STIED, a simple but powerful DL algorithm for the SpatioTemporal detection of focal IEDs. We trained and validated STIED models in a cohort of 10 school-aged children with focal epilepsy (FE) selected based on frequent, high-amplitude, and isolated IEDs arising mainly from the perisylvian areas to which MEG is highly sensitive [22]. These cases thus provide an excellent first case study to benchmark DL models. The sample size was not an issue here as we used a Leave-One-Out Cross-Validation (LOOCV) approach to estimate the performance of the model when used on new patient data. This approach is particularly appropriate for small datasets, especially when model accuracy is more important than the computational cost of training [23]. To test our hypothesis, we compared the performances of three CNN algorithms: a temporal model (1D-CNN applied to a MEG waveform) encoding temporal features only, a spatial model (2D-CNN applied to a magnetic field pattern) encoding spatial features only, and a spatiotemporal model integrating both aspects simultaneously. We show that all models lead to high-performance IED detection metrics (accuracy, specificity, and sensitivity)—with the spatiotemporal model being the most effective—thanks to the combined encoding of IED amplitude with temporal waveform morphology and spatial dipolar magnetic pattern. As a proof of concept, we further tested the generalizability of the spatiotemporal model directly on 12 presurgical patients with refractory focal epilepsy (RFE), i.e., the typical clinical target of MEG diagnostic evaluation in epilepsy.

## II. Methods

### A. Participants and data acquisition

*Participants*: We trained and validated our DL models using a small dataset of 10 children (age: 5-11 years; 5 females and 5 males) suffering from FE, most of them from self-limited epilepsy with centro-temporal spikes (SeLECTS), with a high number of visually detected IEDs (VDS-IED) marked by a clinical magnetoencephalographer (O.F.). The trained

TABLE I
PATIENTS' CLINICAL CHARACTERISTICS

| FE Patient# | AGE (YRS) | SEX | DIAGNOSIS | IED SCALP LOCATION | VDS-IED# |
|---|---|---|---|---|---|
| 1 | 10 | F | Structural Epilepsy | R- CT | 127 |
| 2 | 11 | M | EE-CSWS | R- OT | 15 |
| 3 | 8 | M | SeLECTS | CT- bilateral | 175 |
| 4 | 9 | F | SeLECTS | L- CT | 54 |
| 5 | 10 | M | SeLECTS | L,R-CT; R-O | 68 |
| 6 | 10 | M | SeLECTS | R- CT | 29 |
| 7 | 11 | M | SeLECTS | L- CT | 16 |
| 8 | 5 | F | SeLECTS | L- CT | 97 |
| 9 | 9 | F | SeLECTS | L- CT | 14 |
| 10 | 11 | F | Structural RFE | R- CT | 79 |

| RFE Patient# | AGE (YRS) | SEX | PLEZ | IED LOCATION | |
|---|---|---|---|---|---|
| 11 | 27 | M | L- T | MF (max L-F) | |
| 12 | 15 | F | L- Op | L- Op | |
| 13 | 39 | M | R- OF FCD | R- OF | |
| 14 | 19 | F | L- FT | MF (max L-T) | |
| 15 | 21 | M | R- 1 FCD | R- OI | |
| 16 | 26 | M | R- extratemporal | R- P | |
| 17 | 48 | M | R- F | R- perisylvian | |
| 18 | 27 | M | L- 1 FCD | L- perisylvian; L-T | |
| 19 | 10 | M | L- F | MF (max L-perisylvian) | |
| 20 | 6 | F | L- extratemporal, L-O post-traumatic lesion | L- perisylvian; L-O | |
| 21 | 20 | M | L- OI FCD | L- OI | |
| 22 | 32 | M | L- inferior parietal FCD | L- posterior superior T | |

Abbreviations. FE: focal epilepsy patient number. RFE: refractory focal epilepsy; Patient#: patient number; IED: interictal epileptiform discharge; VDS-IED#: number of visually detected IEDs; PLEZ: presumed location of the epileptogenic zone. F: female; M: male; EE-CSWS: epileptic encephalopathy with continuous spike-and-wave during sleep; SeLECTS: self-limited epilepsy with centro-temporal spikes; R: right; L: left; MF: multifocal; FCD: focal cortical dysplasia; O: occipital; Op: opercular; I: insular; T: temporal; F:frontal; P: parietal; CT: centro-temporal; FT: fronto-temporal; OT: occipito-temporal; OF: orbitofrontal; OI: operculo-insular.

spatiotemporal DL model was also tested on a cohort of 12 patients suffering from RFE, including 3 children (age: 6-15 years; 2 females and 1 male) and 9 adults (age: 19-48 years; 1 female and 8 males). See TABLE I for clinical characteristics of both FE and RFE groups. All FE patients and their legal representative signed a written informed consent after approval by the H.U.B–Hôpital Erasme Ethics Committee. The H.U.B–Hôpital Erasme Ethics Committee and RFE group gave approval for the research use of clinical MEG data acquired in the context of their non-invasive presurgical evaluation.

*Data acquisition*: Participants underwent MEG scanning during a seizure-free wakeful rest of at least 5 minutes in sitting position with their eyes open for the FE group, and one hour of recording in supine position to trigger sleep for the RFE group (see [24] for details on the clinical MEG protocol). Data were recorded using a whole-scalp-covering MEG system (Neuromag Triux; MEGIN, Helsinki, Finland; 0.1–330 Hz band-pass, 1 kHz sampling rate) installed in a lightweight magnetically shielded room (Maxshield, MEGIN; see [25] for details). The 306-channel sensor array combines 102 magnetometers and 204 planar gradiometers. Head movements were tracked by four coils whose position relative to fiducials (nasion and tragi) was digitized beforehand (Fastrack, Polhemus, Colchester, Vermont, USA), along with at least 300 face and scalp points for co-registration with a structural brain 3D T1-weighted magnetic resonance image (MRI) (Intera,

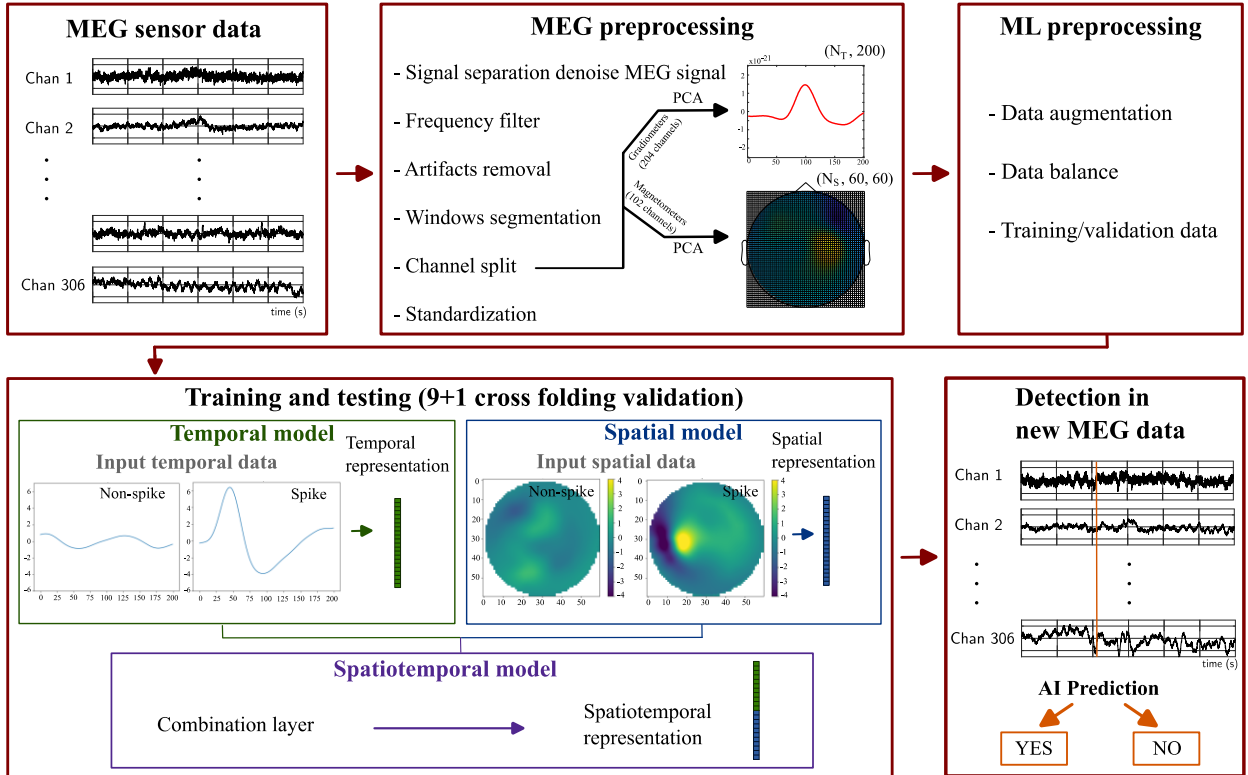

Fig. 1. Schematic representation of the overall deep learning approach. Raw MEG data is MEG- and machine learning (ML-) preprocessed. Training and testing steps are conducted to develop the three DL models (temporal, spatial, spatiotemporal). The detection is performed in new raw MEG data, resulting in a binary yes/no (spike/non-spike) prediction following the LOOCV.

Philips, The Netherlands).

### B. MEG data preparation

The overall pipeline is illustrated in Fig. 1 (upper panel).

*MEG preprocessing:* We followed standard denoising steps of MEG data, i.e., signal space separation to correct for environmental magnetic interferences and head movements (Maxfilter v2.2.14, MEGIN, with default parameters [26]), and ICA (FastICA, [27]) of band-filtered data (0.5–45 Hz) to remove physiological artifacts (eye blinks and cardiac activity) [28]. The resulting data were further band-filtered between 4 and 30 Hz and temporally segmented into 200 ms-long non-overlapping windows. This length was chosen to enable full inclusion of isolated IEDs, whose typical duration range from 50 to 100 ms [29].

*Spatial/temporal feature extraction:* Input data for training/testing of subsequent DL models was extracted from each window by applying principal component analysis (PCA) separately to the 204 planar gradiometer signals and to the 102 magnetometer signals. The first principal component (PC) of gradiometers determined a single waveform per window (size $1 \times 200$ time samples) that explained the largest fraction of variance in the window. The first PC of magnetometers determined a single magnetic topography (size 102 channels $\times$ 1), which was then converted into a 2D image by a fourth order polynomial interpolation over a $60 \times 60$ pixel grid using the open-source software FieldTrip [30]. The splitting of gradiometers/magnetometers to extract temporal/spatial features broadly reproduces the visual IED identification procedure typically followed by our clinical magnetoencephalographers; the higher signal-to-noise ratio of gradiometers yields a cleaner visualization of MEG time courses [10], [31] (notwithstanding lesser sensitivity to deep IED activity in, e.g., mesiotemporal epilepsy [32]) whereas magnetometers show higher sensitivity to dipolar magnetic patterns [33] (as well as to deep IEDs [32]). Extracting a single waveform or image per window also allowed to reduce the size and redundancy of input data, the size and complexity of our DL network models, and the computational load of model training. Since this data reduction could miss potentially relevant, low-variance IED features included in higher-order PCs, we also used second PCs as inputs to DL models and checked whether they bear predictive power for IED detection. Each of these PC data were further standardized per subject.

*Machine learning preprocessing of training data (FE group):* Each window was labeled according to a binary classification ('Spike'/'Non-spike') depending on whether or not it contains an annotated VDS-IED [17] (see Fig. 1, lower left panel). Over the resulting classification dataset, we performed a threefold data augmentation. First, each 'Spike' window was shifted 15 times to the past and future by 7 ms steps, hence allowing to train our DL models independently of the precise IED timing within a window. The number of 30 replicas per 'Spike' window was determined through trial and error. Then, we randomly removed 'Non-spike' windows to achieve a balanced training dataset. Second, we further augmented the dataset spatially by rotating magnetic topographical images four times at 90-degree intervals, so as to train our DL model independently of the precise IED dipole position. Last, we added sign-inverted data given the sign ambiguity of PCA [27]. The final dataset size was 200 time samples × $N_T$ windows for the temporal signal, and $60 \times 60$ pixels × $N_S$ windows for the spatial one.

*IED localization of predicted IEDs (RFE group):* A clinical magnetoencephalographer (X.D.T.) localized brain sources of VDS-IEDs by ECD modeling. For comparison, source localization of IEDs predicted by STIED was performed automatically using noise-standardized Minimum Norm Estimation (see [34] for implementation details) of the highest-amplitude peak within windows containing predicted IEDs, and extracting the location of maximum intensity. To give a sense of clustering of predicted IED localization, we created an IED density map in which each voxel measures the fraction of IEDs predicted at this voxel (i.e., a voxel at 100% concentrates all predicted IEDs, whereas IEDs are spread over large areas in a map with low values).

## C. Model architectures

The architecture of our three DL models is illustrated in Figs. 2-4. All these models were implemented using the Tensorflow library [35] and run on a laptop equipped with an AMD Ryzen 7 5800H processor, 40 GB RAM memory and NVIDIA GeForce RTX 3060 graphic card. Of note, non-trainable hyperparameters were chosen by exploring the space generated by the combination of different number of CNN layers (1, 2, or 3), CNN filters (8, 16, 32, or 64), dense layers (1, 2, or 4), and dense-layer neurons (8, 16, 32, 64).

*Temporal model (1D-CNN):* In the temporal model, the gradiometric PC waveform of a window is passed through three successive blocks made of a 1D convolutional layer followed by dropout and maxpooling layers (Fig. 2). The output was then flattened and fed to a two-layer decision block. Each convolutional layer has a different number of filters (16, 32, 64), a leaky rectified linear unit (ReLU) as activation function, and subsequent dropout layer (probability value 0.5) to prevent overfitting and maxpooling layer (pooling size 2) for downsampling. The decision block was composed of a dense layer with 16 neurons and ReLU activation functions, followed by a single neuron with sigmoid activation function providing a probability between 0 and 1 for a window to contain a spike. The final binary yes/no ('Spike'/'Non-spike') decision was based on a non-trainable probability threshold set to 0.5.

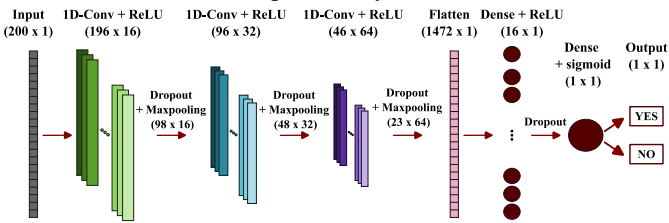

Fig. 2. Temporal model architecture (1D-CNN). The input corresponds to the gradiometric waveform of a time window. The architecture includes three temporal CNNs (1D-Conv, followed by dropout and maxpooling) and a decision block (flattening and two dense layers).

*Spatial model (2D-CNN):* The model design to analyze spatial features was exactly the same as for the temporal model, except that the inputs consisted here in images of magnetometric PC field pattern, so all convolutional layers were 2D (Fig. 3).

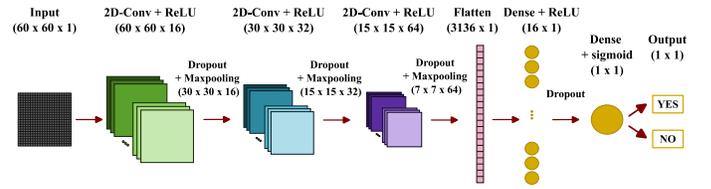

Fig. 3. Spatial model architecture (2D-CNN). The input corresponds to the magnetometric field topography of a time window. The architecture includes three spatial CNNs (2D-Conv, followed by dropout and maxpooling) and a decision block (flattening, and two dense layers).

*Spatiotemporal model (STIED):* To incorporate both temporal and spatial features in a genuinely spatiotemporal IED detection, we designed a model architecture that processes temporal and spatial inputs in parallel using the same 1D-CNN (Fig. 2) and 2D-CNN (Fig. 3) up to the first dense layer of their decision block. The feature vectors from the outputs of the temporal and spatial dense layers were downsampled (0.5 dropout), combined using weighted concatenation with one trainable scalar weight for the temporal features ($w_t$) and one for the spatial features ($w_s = 1 - w_t$), and finally fed to the same decision block as above (Fig. 4).

*Amplitude normalized models and threshold models:* Since IEDs are primarily high-amplitude events, we sought to examine whether our STIED models are merely driven by the gross amplitude of IEDs or if they learn subtler features. To design a classifier independent of amplitude, we applied the same 1D-CNN, 2D-CNN, and STIED models on input data after normalizing their peak amplitude to one within each window (NormAmplitude models). The specific importance of amplitude, independently of IED morphological features, was assessed using a classical DL classification model (multilayer perceptron with two fully connected dense layers of 8 neurons each and ReLU activation function) determining a trainable threshold for the scalar peak amplitude of each input data (ThresholdAmplitude model).

## D. Training and performance evaluation

*Model training:* Model training was performed using an Adam optimizer algorithm with a batch size of 128 windows, binary cross-entropy as loss function, and a reduced learning rate when loss stopped improving. As preliminary sanity check, we trained our STIED models by randomly extracting 80% of the entire dataset for training and using the remaining 20% for testing. We did not observe overfitting and determined that 150 epochs/training steps were enough to reach a learning plateau in all models. For model performance evaluation, given that our dataset was composed of 10 patients, we performed a 10-fold repeated LOOCV wherein all windows of 9 out of 10 subjects were used for training, and all windows of the remaining subject for testing. This allowed to evaluate model performance in a clinically oriented setup (i.e., generalizability to the inclusion of new patients with similar epilepsy) and inter-individual variability of performance metrics.

*Performance metrics:* For each cross-validation iterate, we compared model window classification (yes/no) to the actual presence of VDS-IEDs ('Spikes'/'Non-Spike') using accuracy, specificity, and sensitivity. We evaluated these metrics using balanced training data (equal number of 'Spikes'/'Non-Spike' windows) as done in previous works ([18], [20], [21]), but also using the full, unbalanced training data as this corresponds to the realistic case of diagnosing a new patient. Statistical comparisons of performance metrics across models were performed using ANOVA and *t* tests.

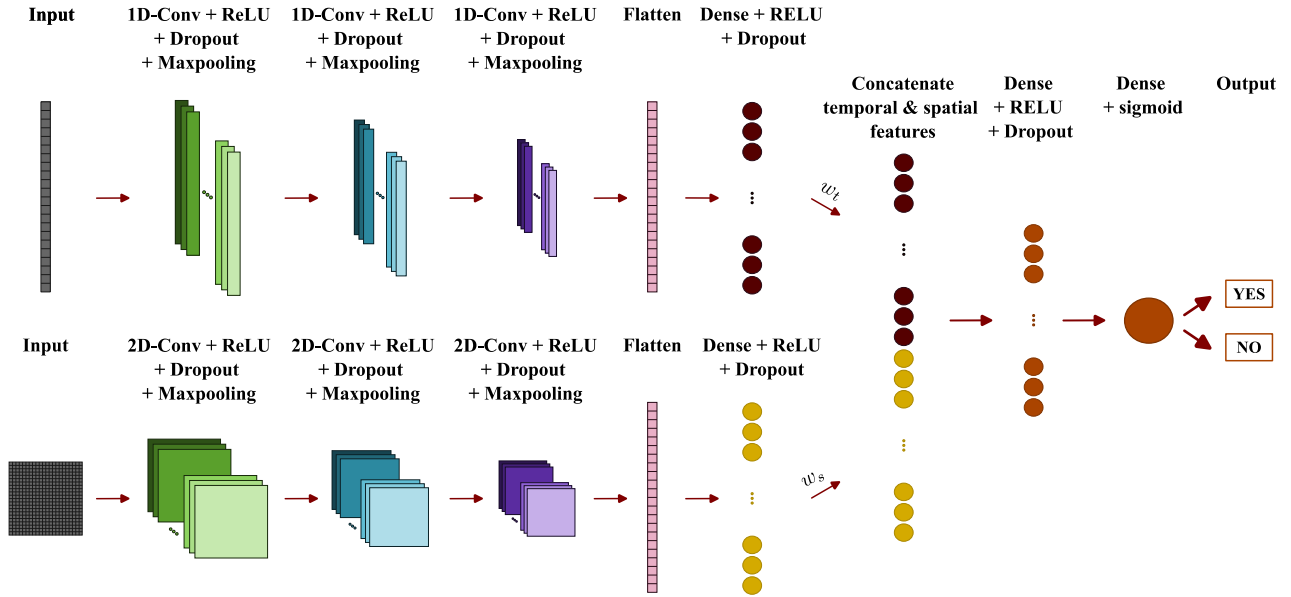

Fig. 4. Spatiotemporal model (STIED): The input corresponds to both the gradiometric waveform and the magnetometric topography of a time window. The architecture incorporates both 1D-CNN and 2D-CNN up to their dense layer, an extra dropout, a weighted concatenation with trainable weights ($w_t$ and $w_s$), and a common decision block (two dense layers).

## III. RESULTS

We now report the detection performance of these supervised DL models and analyze the impact of using temporal, spatial, or spatiotemporal features. We also explore the validity of our data reduction, assess the relevance of IED amplitude and IED morphology in STIED, and compare it with previous unsupervised detection algorithms. Then, we test the generalizability of STIED to another, unlabeled dataset of patients with RFE.

### A. Detection performance of STIED in SeLECTS

The performance metrics for the three DL model types (temporal 1D-CNN, spatial 2D-CNN, spatiotemporal STIED) are summarized in TABLE II for balanced and unbalanced test data. Results at each cross-validation iterate are detailed in TABLE III for unbalanced test data.

*Impact of model type:* Classification accuracy and specificity were consistently large whatever the model type, both for balanced and unbalanced test data (accuracy >85%, specificity >93%; mean over 10 cross-validation iterates), with no significant effect of model type (p>0.12). So, all three DL models were equivalent in their ability to correctly predict VDS-IEDs. Sensitivity was on average larger for the temporal and spatiotemporal models (>84%) than for the spatial model (75%), though this observation was not statistically significant due to large inter-subject variability in sensitivity (p>0.26). This suggests that temporal features of IED waveforms are most useful to avoid false detections. Despite the absence of statistical effects, combining spatiotemporal features maximized all three performance metrics in the realistic setup of unbalanced test data (TABLE II), indicating a slight but possible superiority of the STIED model, on which we shall focus hereafter. Interestingly, the trained weights $w_t$ and $w_s$ were both close to 0.5, showing that while STIED did not perform significantly better than the temporal 1D-CNN model, it effectively used both temporal and spatial features in its decision process.

*Balanced vs. unbalanced metrics:* The observation of similar performance metrics from balanced and unbalanced test data (TABLE II) further advocates for the robustness of our DL algorithms. Specifically, the identical sensitivities (combined with consistently large specificities) indicate that all three models identified the same IEDs regardless of the amount of

TABLE II
10-FOLD LOOCV DETECTION PERFORMANCE FOR A BALANCED AND UNBALANCED DATASET (MEAN ± STANDARD DEVIATION %)

| Model\Metric | TEMPORAL | | | SPATIAL | | | SPATIOTEMPORAL | | |
|---|---|---|---|---|---|---|---|---|---|
| | ACCURACY | SPECIFICITY | SENSITIVITY | ACCURACY | SPECIFICITY | SENSITIVITY | ACCURACY | SPECIFICITY | SENSITIVITY |
| Balanced | 90 ± 5 | 95 ± 6 | 84 ± 13 | 85 ± 9 | 96 ± 3 | 75 ± 19 | 89 ± 8 | 93 ± 4 | 85 ± 14 |
| Unbalanced | 92 ± 4 | 93 ± 5 | 84 ± 13 | 95 ± 3 | 96 ± 2 | 75 ± 19 | 95 ± 2 | 95 ± 2 | 85 ± 14 |

| Patient | TEMPORAL | | | SPATIAL | | | SPATIOTEMPORAL | | |
|---|---|---|---|---|---|---|---|---|---|
| | ACC | SPE | SEN | ACC | SPE | SEN | ACC | SPE | SEN |
| 1 | 92 | 93 | 83 | 95 | 96 | 77 | 93 | 93 | 90 |
| 2 | 92 | 92 | 67 | 91 | 91 | 80 | 93 | 94 | 80 |
| 3 | 98 | 98 | 91 | 97 | 97 | 93 | 97 | 98 | 87 |
| 4 | 90 | 90 | 87 | 97 | 98 | 87 | 96 | 96 | 94 |
| 5 | 93 | 93 | 96 | 96 | 98 | 59 | 95 | 96 | 79 |
| 6 | 89 | 89 | 97 | 97 | 99 | 34 | 95 | 96 | 52 |
| 7 | 98 | 98 | 88 | 97 | 98 | 81 | 97 | 97 | 100 |
| 8 | 93 | 95 | 75 | 92 | 93 | 78 | 91 | 91 | 91 |
| 9 | 83 | 82 | 100 | 96 | 96 | 100 | 96 | 96 | 100 |
| 10 | 89 | 99 | 63 | 87 | 97 | 58 | 92 | 96 | 80 |
| **Mean** | **92** | **93** | **84** | **95** | **96** | **75** | **95** | **95** | **85** |
| **± SD** | **± 4** | **± 5** | **± 13** | **± 3** | **± 2** | **± 19** | **± 2** | **± 2** | **± 14** |

Abbreviations. LOOCV: Leave-One Out Cross-Validation. ACC: Accuracy. SPE: Specificity. SEN: Sensitivity. SD: Standard Deviation.

data without VDS-IEDs. In other words, DL model performance did not depend on IED frequency of the test patient. This conclusion was further supported by the absence of correlations between performance metrics and number of VDS-IEDs (Spearman correlation, $|r <0.09$; $p>0.36$).

*Impact of data reduction:* The first PC used for data reduction explained significantly more data variance than the second PC of same windowed data ($t$ test on window covariance eigenvalues, $p<3e-6$), but this does not preclude that the latter contains useful IED features. Figure 5 compares performance metrics (unbalanced test data) for our three DL models applied to the first vs. second PC. All metrics were significantly smaller when using the second PC in STIED ($p<0.01$), with a particularly strong drop in mean sensitivity of 52%. The drop in accuracy and sensitivity were similarly significant in the temporal and spatial models ($p<0.05$) but not in specificity ($p>0.11$). Thus, the second PC contains remnant, low-variance features of IEDs excluded by our data reduction procedure, though these features are less effective at IED detection and are poorly specific to IEDs.

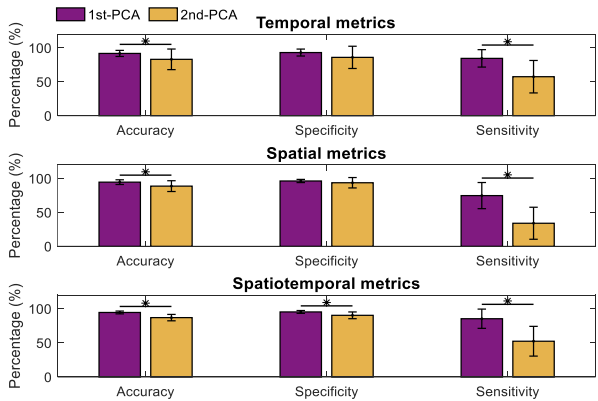

Fig. 5. Detection performance of the temporal, spatial and spatiotemporal models obtained from 1st-PC vs. 2nd-PC (unbalanced test data). Bar plots show mean values across 10-fold cross-validation iterates, and vertical bars show standard errors. "*" indicates statistical significance ($t$ tests).

*Contribution of IED amplitude:* Figure 6 compares our DL models (Full) to similar models wherein the encoding of IED amplitude is explicitly precluded (NormAmplitude), and to a simple amplitude threshold model that solely encodes IED amplitude (ThresholdAmplitude). Remarkably, eliminating the amplitude feature did not alter performance metrics (unbalanced test data) for the three model types, except for a significant decrease of accuracy/specificity ($p<0.01$; NormAmplitude vs. Full) in the spatial model (2D-CNN) and a marginal (non-significant; $p=0.09$) decrease of sensitivity in the spatiotemporal model (STIED). Amplitude-based classification led to significantly lower accuracy/specificity ($p<0.004$; except for spatial ThresholdAmplitude vs. NormAmplitude where $p>0.09$) but higher sensitivity, though the latter effect was marginal in the original STIED model ($p>0.25$). We conclude that IED amplitude is not effective in recognizing IEDs and plays a relatively marginal role in STIED. In other words, STIED is mostly driven by IED waveform morphology and dipolar topography.

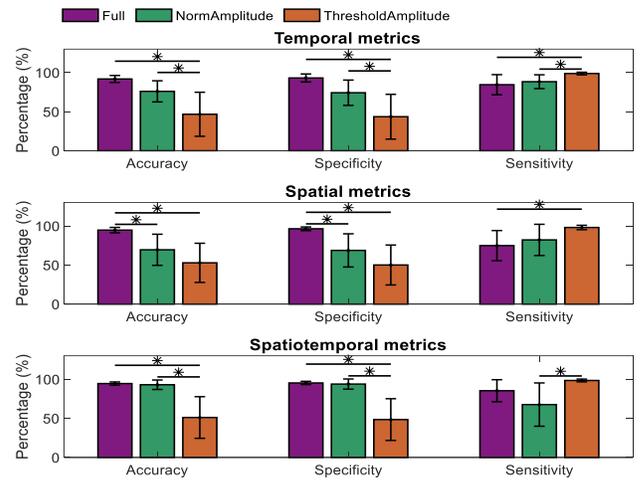

Fig. 6. Detection performance metrics (unbalanced test data) of the three DL *Full* models (purple; top, 1D-CNN; middle, 2D-CNN; bottom, STIED) compared with *NormAmplitude* models encoding morphological and/or topographical IED features independently of their amplitude (green), and with *ThresholdAmplitude* models encoding the amplitude feature only (orange). Bar plots show mean values across 10-fold LOOCV, and vertical bars show standard errors. "*" indicates statistical significance (ANOVA, post-hoc $t$ tests).

*Comparison with unsupervised classifications:* TABLE IV compares performance metrics of STIED (unbalanced test data) to those of two unsupervised detection algorithms (ICA and HMM) developed in a previous work and tested on the same dataset [17]. All approaches showed similar sensitivities ($p=0.67$), but STIED was significantly more specific to IEDs than both supervised methods ($<55\%$; $p<2e-9$). This merely reflects the expected fact that the supervised DL algorithm learns from VDS-IEDs and thus follows closely the clinical magnetoencephalographer's decision.

TABLE IV
SUPERVISED VS. UNSUPERVISED DETECTION PERFORMANCE

| Metric (mean ± SD) \ Model | STIED | ICA | HMM |
|---|---|---|---|
| Sensitivity | 85 ± 14 | 91 ± 13 | 85 ± 19 |
| Specificity | 95 ± 2 | 52 ± 2 | 55 ± 4 |

Abbreviations. STIED: Spatiotemporal interictal epileptiform discharges. ICA: Independent Component Analysis; HMM: Hidden Markov Modeling. SD: Standard Deviation.

## B. Generalizability of FE-group-trained STIED to RFE patients

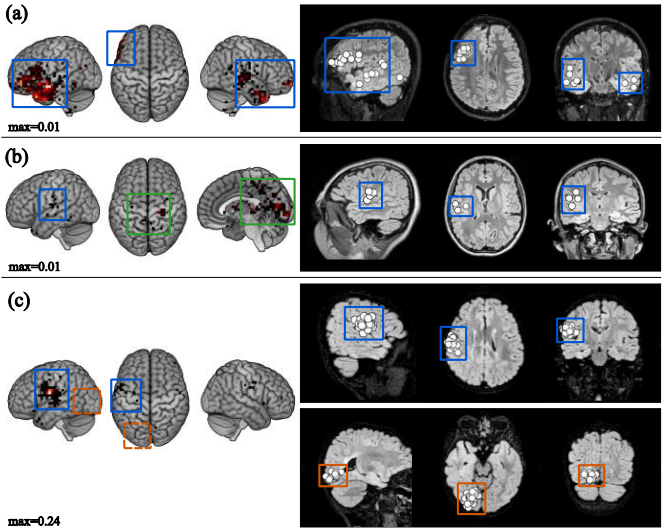

Fig. 7. Density map of IEDs predicted by STIED and ECD localization of VDS-IEDs. (a) Successful example of coincident localization (blue squares) (patient 11). (b) Partially successful example where non-epileptiform events (green squares) are detected, such as Rolandic and sleep physiological transients (patient 12) (c) Mitigated example where a cluster of VDS-IEDs is missed (orange squares) in the left-occipital part (patient 20). All MRIs are in neurological convention.

As a final step, we sought to evaluate the performance of our STIED model, trained in our FE group, on 12 patients with RFE. Besides differences in the type of epilepsy, this test dataset differs from the training dataset in that (i) it contains both children and adults, (ii) patients slept during MEG recording, and (iii) IEDs were not marked, though their ECD location was available, allowing comparison of localization with STIED detection. Perphaps unsurprisingly, results were mitigated with only 2 patients reproducing exactly the location of IED events, 9 patients in which additional non-epileptiform events were detected besides the accurate location of VDS-IEDs, and 1 patient in which a cluster of VDS-IEDs was missed altogether.

Figure 7 illustrates these results by comparing the density map of predicted IEDs (Fig. 7, left) and ECD localization of VDS-IEDs (Fig. 7, right) in three patients. Patient 11 (Fig. 7a) is representative of the successful cases where STIED co-localizes IEDs with VDS-IEDs. Both approaches were consistent with multifocal independent irritative zones, with the most active being posterior to the left frontal lesion. Patient 12 (Fig. 7b) illustrates a case of partial success. Clinical analysis reported few VDS-IEDs, which were accurately identified by the STIED model. Nonetheless, Rolandic spikes in the centrotemporal area and sleep-related events in the occipital areas were also wrongly predicted, likely indicative of physiological transients [10], [36]. Patient 20 (Fig. 7c) illustrates the case of mitigated success. Clinical assessment revealed two active, independent irritative zones located at the left opercular/periinsular and ventral occipital regions, which might be part of a common epileptogenic network, but STIED missed the ventral occipital cluster of VDS-IEDs. We conclude that, while STIED trained on FE patients learned IED features that generalize to all patients with RFE, other types of epileptiform activities were not recognized (such as polymorphic waveform), while some forms of physiological activity were falsely detected.

## IV. DISCUSSION

The development of fully automated tools for the detection and localization of IEDs is a long-standing goal in the field of clinical epilepsy. Here, we introduced STIED, a simple yet effective DL solution for IED automatic detection in clinical MEG recordings. This is part of the global momentum that "artificial intelligence" bears in technology, not the least in medical fields—including neuroscience [37], [38] and clinical epilepsy [39]. In fact, STIED is not the first attempt in applying supervised DL to the evaluation of epilepsy with MEG ([18]–[21]). Here, we strived to design a computationally reasonable architecture that encodes both temporal and spatial features in a way analogous to how clinical magnetoencephalographers analyze their clinical MEG recordings [10]. This allowed us to train our STIED model with high sensitivity and specificity, despite a relatively small training dataset, at least in the context of our FE group. We focused here on patients with FE, mainly SeLECTS, because their high IED amplitude and frequency yielded a relatively large count of IED events in a limited duration of MEG signals, which additionally may have contributed to the efficient training of STIED at small dataset size. Interestingly, our results suggest that, after training, detection performance does not depend closely on IED frequency and thus could generalize to epileptic patients with scarcer interictal events. This contrasts with previous recent studies [18]–[20] proposing DL-based IED detection, which showed a substantial drop in sensitivity to less than 50% when assessed with unbalanced compared to balanced test data [20]. While authors overcame this issue *ad-hoc* by adjusting the probability threshold at the end of their decision blocks (using a thresholding moving strategy to enhance performance [18]– [20]), one advantage of our DL models is that they showed high sensitivity without such adjustment. Still, the generalizability of our current version of STIED to other types of epilepsy such as RFE was partial, presumably because it does not capture sufficiently the variety of IED field patterns and waveform morphologies (which may be different in deep medial temporal epilepsy [40] or change as the type of epilepsy evolves with age [41]). This being said, this paper provides proof of concept for the effectiveness of the STIED model architecture. Future updates of STIED with a larger and more diverse training dataset should dramatically widen its range of applicability. STIED could also be an efficient and fast approach to assess the spike-wave index (i.e., fraction of recording time showing spike-and-wave activity [42]), a key diagnostic element for some pediatric epileptic conditions (reaching up to 85% in our training dataset [43]).

The STIED architecture is the result of several trials not reported here. Regarding artificial neural network design, we quickly settled for a quite classical setup that combines 1D-CNN as done in previous works to analyze temporal signals [44], 2D-CNN typical of spatial image processing [45], and dense layers with sigmoid final neuron for classification. The number and size of layers was varied to maximize performance while minimizing complexity. In this endeavor, the identification of CNN inputs with relevant low-dimensional features amongst high-dimensional MEG data, proved key. We did so using a data reduction based on PCA that loosely follows what is done during clinical MEG visual analysis, i.e., extract IED waveform signals from gradiometers and images of IED dipolar patterns from magnetometers. This likely explains why STIED could learn from, and closely agree with, clinical magnetoencephalographers. A caveat of this procedure was the exclusion of MEG features specific of, though poorly sensitive to, IEDs. It might also miss deep epileptiform activity [32]. On the other hand, it allowed to train and run the model on affordable hardware with reasonable training times (about 4 hours for 10 subjects with 5-min MEG recordings) and small processing time of new data (less than 30 seconds for 1 hour MEG recording). This means that STIED can detect IEDs similarly to clinical magnetoencephalographers, about 300 time faster—again at least within our FE group.

Beyond sheer performance, we sought to partially reverse engineer the STIED model and investigate what features of MEG signals drive IED detection. A surprising finding is that the amplitude of IED waveforms— at first sight a primary gross characteristic of epileptiform activity—was not critical in the STIED decision process. In fact, IED amplitude alone was poorly specific of visually identified epileptiform activity. Interestingly, a similarly low specificity was found in unsupervised algorithms based on ICA and HMM, for which arguments were given for the hypothesis that false positive events actually correspond to genuine but small-amplitude IEDs unidentified by clinical MEG evaluation [17]. This illustrates a possible advantage of unsupervised models. On the other hand, STIED yielded a fully automated decision, whereas both ICA and HMM were semi-automated. Further, whether such small-amplitude events, unidentified by STIED too, represent IEDs or clinically irrelevant physiological events, remains to be fully clarified. In the current state of knowledge, a conservative approach may be warranted for clinical applications so STIED likely represents the best data-driven approach so far. In the future, it will be interesting to compare the outcome of STIED and unsupervised algorithms in other types of IED recordings, such as the ICA of MEG based on optically pumped magnetometers [46] or the HMM of stereo-electroencephalography data [47].

Our reverse engineering analysis also revealed that, while STIED does rely on both its spatial and temporal inputs (as we intended), the spatial features of IED magnetic topography bring at most a subtle added value to detection performance, compared with the temporal morphology of IEDs. In hindsight, this is likely because IED spiking waveforms are automatically accompanied with dipolar magnetic field patterns, so the latter do not add independent information to the former. Indeed, epileptiform spikes originate from bursts of hyperexcitable neurons within focal neural populations. In neocortical epilepsies (as was the case in our training group), focal excitation leads to post-synaptic current flows along apical dendrites of pyramidal neurons that are necessarily dipolar [48]. More generally, all sharp electrophysiological events—not only IEDs but also physiological transients typically occurring in Rolandic, supramarginal, and occipital areas [10], [36] as well as transient oscillatory bursts emerging in brain functional networks [49]–[51]—are focal and thus dipolar. Spatial features thus hardly distinguish sharp physiological events from IEDs, which presumably explains the poorer performance of the spatial DL model. Another, more technical implication is that data augmentation by rotation of topographical patterns should not be overdone. While we initially reasoned (following standard image recognition processing) that such rotations would improve model generalizability across patients by allowing more diverse IED localizations, during our trial phase we found that coarse 90-degree rotations used here worked better than finer rotations (e.g., at 12 degrees, the sensitivity of the spatial DL model tended to increase slightly, but specificity dropped significantly, $p < 0.03$). Associating IEDs with too many dipolar patterns presumably renders DL models oversensitive to sharp physiological events. This aspect will require further consideration when updating STIED with a larger training dataset. On the other hand, including spatial features might help against false detections of spike-like artefacts that may arise in MEG signals with atypical dipolar patterns.

The problem of physiological confounds was somewhat alleviated by the inclusion of temporal features, yet physiological transients with spiking waveform morphology remain a critical challenge to overcome [28] as they are detrimental to detection specificity. This was illustrated by the high rate of false detections, together with the accurate location of VDS-IEDs, when applying our current version of STIED to sleep recordings of 9 out 12 patients with RFE. Some physiological transients typically emerge during drowsiness and sleep [28], so a critical next step might be to include sleep recordings of healthy controls to help STIED learn to distinguish them from epileptiform activity.

## V. CONCLUSION

In sum, we provided proof of concept that STIED is a promising DL solution for the automated detection of IEDs in focal epilepsy. In future work, we intend to update the current version with a large-scale training dataset of labelled MEG recordings in patients with different types of epilepsies, including sleep recordings to discriminate physiological spikes. If expected improvements in model generalizability and detection specificity are confirmed, this would establish STIED as an invaluable tool to assist clinical magnetoencephalographers in getting a fast and accurate clinical evaluation of epilepsy from MEG recordings.

## AUTHOR CONTRIBUTION

Raquel Fernández-Martín (R.F.M.) and Alfonso Gijón (A.G.) contributed to the conceptualization, methodology, design and implementation of the algorithm, software, formal analysis, investigation, validation, visualization and writing (original draft, review and editing); Odile Feys (O.F.) contributed to data acquisition, clinical analysis (localization of IEDs) and writing (review); Elodie Juvené (E.J.) contributed to data acquisition and writing (review); Alec Aeby (A.A.) contributed to writing (review) and resources; Charline Urbain (C.U.) contributed to writing (review) and resources; Xavier De Tiège (X.D.T.) contributed to conceptualization, data acquisition and clinical data analysis (localization of IEDs), supervision, writing (review), resources and funding; Vincent Wens (V.W.) contributed to conceptualization, methodology, investigation, validation, supervision and writing (original draft, review and editing).